# Traffic flow on star graph: Nonlinear diffusion

Takashi Nagatani*

*Department of Mechanical Engineering, Shizuoka University, Hamamatsu 432-8561, Japan*

### Abstract

We study the urban-scale macroscopic traffic flow in city networks. Star graph is considered as traffic network. Star graphs with controlled traffic flow are transformed to various cell-transmission graphs by using the cell transmission method. The dynamic equations of vehicular densities on all nodes (roads) are presented on cell-transmission graphs by using the speed-matching model. The density equations are given by nonlinear-diffusion equations. The traffic flow on star graph is mapped to the nonlinear diffusion process on the cell-transmission graphs. At low mean density, the dynamic equations of densities can be approximated by the conventional diffusion equations. At low and high mean densities, the analytical solutions of densities on all nodes (roads) are obtained on cell-transmission complete, cycle and star graphs. By solving the dynamic equations numerically, the densities on all roads are derived at a steady state. The urban-scale macroscopic fundamental diagrams are obtained numerically on the cell-transmission graphs. The analytical solutions agree with the numerical solutions.

*Email address: nagatani.takashi@shizuoka.ac.jp, wadokeioru@yahoo.co.jp

# 1. Introduction

Transportation problems have been investigated extensively in engineering, physics, mathematics, and interdisciplinary research fields. The concepts and techniques of physics and mathematics have been applied to transportation systems. The traffic dynamics has been studied from a point of view of statistical mechanics and nonlinear dynamics [1-6]. Especially, traffic jam, congestion, signal control, and route choice have been investigated numerically and analytically [7-27].

Recently, complexity science and information systems in networks have attracted great attention [28,29]. Most modeling and analysis techniques are presented to study dynamic processes in static, temporal, and adaptive network [30-33]. The dynamic modelling has been done in biological, social, and financial networks.

Recently, urban-scale transportation system has drawn great attention for traffic flow in city networks [34-39]. In urban-scale transportation, traffic network is formed in a city because roadways are connected each other. The urban-scale macroscopic fundamental diagram can make a rough prediction of urban-scale traffic. The urban-scale macroscopic fundamental diagram has not been derived successfully for extended densities except for low density by using the conventional macroscopic model because of traffic instability. A new macroscopic traffic model without instability has been presented by matching speed at downstream. The speed-matching model for the traffic flow in the network does not induce the traffic instability even if traffic congestion occurs. The macroscopic fundamental diagram has been reproduced successfully by numerical simulation [40].

However, the macroscopic fundamental diagram has not been derived analytically. The effect of network structure on the traffic flow remains poorly understood. It is necessary and important to present the traffic model on simple networks for obtaining an analytical solution. How does the network structure affect the traffic flow? How does the urban-scale macroscopic fundamental diagram change by the network structure? In network modelling, star graph is simple and specific because it has a hub with high degree. We consider the traffic flow on star graph for obtaining analytical solutions. On star graph, traffic flow is controlled by restricting outflow from a loop (road). Various cell-transmission graphs are realized by controlled traffic flow on star graph.

In this paper, we study the urban-scale traffic flow in networks. We transform the controlled traffic flow on star graph to that on cell-transmission graphs. We present a dynamic model of vehicular densities for the traffic flow on cell-transmission graphs. The dynamic model produces the traffic flow stabilized by matching speed. We map the traffic flow on star graph to the nonlinear diffusion process on the cell-transmission graphs. At low density, the dynamic equations are approximated by the conventional diffusion equations. Also, the dynamic equations are given at high density. We derive the analytical solutions of vehicular densities in a steady state at low and high densities. We carry out numerical simulation and obtain the numerical solutions. We compare the analytical result with the numerical result.

## 2. From traffic star graph to cell-transmission graphs

We study the traffic flow on star graph. The traffic flow is controlled when vehicles move out from a loop, through the intersection, to other loops. We consider the star graph in Fig. 1(a). Figure 1(a) shows the schematic illustration of the star graph with $N = 3$ where $N$ is the number of loops. Each loop represents one road. Loops 1, 2, and 3 are indicated by black, red, and blue colors. There is an intersection at the center. Vehicles move uni-directionally on each loop. After vehicles go round on a loop, they move out of the loop. After vehicles move out of loop 1, they are divided into two equal parts and flow into loops 2 and 3. Similarly, after they move out of loop 2, vehicles flow into loops 1 and 3 one half each. After vehicles move out of loop 3, they flow into loops 1 and 2 one half each. The process is repeated. The traffic flow from loop 1 (2, 3) to other loops is indicated by black (red, blue) arrow.

In traffic networks, nodes and loops correspond to intersections (junctions) and roads (streets) respectively. We transform the traffic network to the simple graph by using the cell transmission method proposed by Daganzo [38,39]. The traffic network (a) is transformed to the graph (b) in Fig. 1. In the transformed graph, each node defines a road segment. Loops 1, 2, and 3 are represented by black, red, and blue nodes. A link represents the connectivity between roads. Traffic flow from node 1 (2, 3) to other nodes is represented by black (red, blue) arrow. In this case, the transformed graph is the complete graph with $N = 3$ where $N$ is the number of nodes. The transformed graph has bi-directional links. We call the transformed graph as the cell-transmission graph. We note that the directionality is important in traffic flow and traffic flow from node 1 to node 2 is different from that from node 2 to node 1.

Next, we consider the star graph in Fig. 1(c). Figure 1(c) shows the schematic illustration of the traffic flow different from that in Fig. 1(a). After vehicles move out of loop 1, they are divided into two equal parts and flow into loops 2 and 3. After vehicles move out of loop 2, they flow only into loop 1.

After vehicles move out of loop 3, they flow only into loop 1. The process is repeated. The traffic flow from loop 1 (2, 3) to other loops is indicated by black (red, blue) arrow. We transform the traffic network to the simple graph by using the cell transmission method. The traffic network (c) is transformed to the graph (d) in Fig. 1. Traffic flow from node 1 (2, 3) to other nodes is represented by black (red, blue) arrow. In this case, the transformed graph is the cell-transmission star graph with $N=3$. There is no links between nodes 2 and 3.

Similarly, we obtain the cell-transmission graphs for traffic star graph with $N=4$. In the cell-transmission graphs, indegree $k_{in}$ is consistent with outdegree $k_{out}$, $k_{in}=k_{out}$. Figure 2 shows all cell-transmission graphs obtained from the traffic star graph with $N=4$. Transformed graph (a) is cell-transmission complete graph. Transformed graph (b) is cell-transmission cycle graph. Graphs (a) and (b) are homogeneous. Transformed graph (c) is cell-transmission star graph. Graph (f) is cell-transmission path (chain) graph. Graphs (c)-(f) are heterogeneous. Thus, the traffic star graphs with controlled flow can be transformed to cell-transmission graphs by the cell-transmission method. We study the traffic behavior on the cell-transmission graphs.

## 3. Density equations on cell-transmission graphs

We derive the density equations of traffic flow on cell-transmission graphs with $k_{in}=k_{out}$. The density on node $i$ at time $t$ is defined as $\rho_i(t)$. The length of road (loop) is $l_i$. The road length is assigned to each node. The number of vehicles existing on node $i$ is $l_i\rho_i(t)$. The road length is integrated as the capacity of nodes on the cell-transmission graph. The total number $M$ of vehicles on the traffic network is conserved and constant:

$$M = \sum_{i=1}^{N} l_i \rho_i(t), \tag{1}$$

where $N$ is the number of nodes (roads) on the cell-transmission graph. The outflow depends not only on the density at the road but also on the density at the downstream. On the directed graph, the traffic inflow from node $j$ to node $i$ is $J_{j,i}(\rho_j, \rho_i)$. The traffic outflow from node $i$ to node $j$ is $J_{i,j}(\rho_i, \rho_j)$. Generally, inflow $J_{j,i}(\rho_j, \rho_i)$ is not equivalent to outflow $J_{i,j}(\rho_i, \rho_j)$. The dynamical equation of densities on node $i$ is given by

$$l_i \frac{d\rho_i}{dt} = \sum_{j=1}^{k_i} J_{j,i}(\rho_j, \rho_i) - \sum_{j=1}^{k_j} J_{i,j}(\rho_i, \rho_j), \tag{2}$$

where $k_i$ is the degree of node $i$, $k_i = k_{in} = k_{out}$.

The inflow from node $j$ to node $i$ depends on both densities at nodes $j$ and $i$. When a congestion occurs on the downstream, the traffic inflow to the downstream reduces greatly and the vehicular speed at the downstream is rather low. The effect is taken into account as follows. Drivers try to match their speed to that on the downstream. The modelling is called the speed-matching model [40]. The inflow from node $j$ to node $i$ and outflow from node $i$ to node $j$ are given by

$$\begin{aligned} J_{j,i}(\rho_j, \rho_i) &= \rho_j(t) u_i(t) / k_j, \\ J_{i,j}(\rho_i, \rho_j) &= \rho_i(t) u_j(t) / k_i, \end{aligned} \tag{3}$$

where $u_i(t)$ is the vehicular velocity at downstream node $i$.

Density equation (2) with Eq. (3) is a kind of nonlinear diffusion equations with diffusion coefficient $u_i(t)$. Thus, the traffic flow on star graph is mapped to the nonlinear diffusion process on the cell-transmission graphs.

Here, we assume that the velocity on node $i$ is given by the following obtained from the so-called fundamental diagram on a single road:

$$\begin{aligned} u_i(\rho_i) &= v_{max} && \text{if } \rho_i \leq 1/(v_{max} + 1), \\ u_i(\rho_i) &= (1 - \rho_i) / \rho_i && \text{if } \rho_i > 1/(v_{max} + 1), \end{aligned} \tag{4}$$

where $v_{max}$ is the maximal velocity. When $v_{max}=1$, Eq.(4) is the velocity in CA 184 model [2]. When $v_{max}$ is an integer higher than one, Eq. (4) is the velocity in Fukui-Ishibashi CA model [2].

At low density of $\rho_i \leq 1/(v_{max}+1)$, the density equation (2) with Eq. (3) reduces the conventional diffusion equation with diffusion constant $v_{max}$ (see Sec. 4).

We apply density equation (2) to the traffic flow on cell-transmission graph in Fig. 1(b). The density equations at nodes 1, 2, and 3 are given by

$$l_1 \frac{d\rho_1}{dt} = J_{2,1}(\rho_2,\rho_1) + J_{3,1}(\rho_3,\rho_1) - J_{1,2}(\rho_1,\rho_2) - J_{1,3}(\rho_1,\rho_3),  \qquad (5)$$

$$l_2 \frac{d\rho_2}{dt} = J_{1,2}(\rho_1,\rho_2) + J_{3,2}(\rho_3,\rho_2) - J_{2,1}(\rho_2,\rho_1) - J_{2,3}(\rho_2,\rho_3),  \qquad (6)$$

$$l_3 \frac{d\rho_3}{dt} = J_{1,3}(\rho_1,\rho_3) + J_{2,3}(\rho_2,\rho_3) - J_{3,1}(\rho_3,\rho_1) - J_{3,2}(\rho_3,\rho_2),  \qquad (7)$$

with

$J_{1,2}(\rho_1,\rho_2) = \rho_1 u_2(\rho_2)/2$, $J_{1,3}(\rho_1,\rho_3) = \rho_1 u_3(\rho_3)/2$, $J_{2,1}(\rho_2,\rho_1) = \rho_2 u_1(\rho_1)/2$, $J_{3,1}(\rho_3,\rho_1) = \rho_3 u_1(\rho_1)/2$, $J_{2,3}(\rho_2,\rho_3) = \rho_2 u_3(\rho_3)/2$, $J_{3,2}(\rho_3,\rho_2) = \rho_3 u_2(\rho_2)/2$.

We apply density equation (2) to the traffic flow on cell-transmission graph in Fig. 1(d). The density equations at nodes 1, 2, and 3 are given by

$$l_1 \frac{d\rho_1}{dt} = J_{2,1}(\rho_2,\rho_1) + J_{3,1}(\rho_3,\rho_1) - J_{1,2}(\rho_1,\rho_2) - J_{1,3}(\rho_1,\rho_3),  \qquad (8)$$

$$l_2 \frac{d\rho_2}{dt} = J_{1,2}(\rho_1,\rho_2) - J_{2,1}(\rho_2,\rho_1),  \qquad (9)$$

$$l_3 \frac{d\rho_3}{dt} = J_{1,3}(\rho_1,\rho_3) - J_{3,1}(\rho_3,\rho_1),  \qquad (10)$$

with

$J_{1,2}(\rho_1,\rho_2) = \rho_1 u_2(\rho_2)/2$, $J_{1,3}(\rho_1,\rho_3) = \rho_1 u_3(\rho_3)/2$, $J_{2,1}(\rho_2,\rho_1) = \rho_2 u_1(\rho_1)$, $J_{3,1}(\rho_3,\rho_1) = \rho_3 u_1(\rho_1)$.

Similarly, we obtain the density equations at all nodes for the traffic flow on cell-transmission graphs with $N = 4$ in Fig. 2.

## 4. Conventional diffusion approximation

We approximate the density equation (2) in terms of conventional diffusion equation for low-density traffic. At low density, traffic current is proportional to the density from Eq. (4). The proportional coefficient is given by maximal velocity $v_{max}$. The vehicles flowing out from node $i$ divides into $k_i$ parts and flows into node $j$. The traffic current from node $i$ to node $j$ is approximated by

$$J_{i,j}(\rho_i, \rho_j) = \rho_i(t)u_j(t)/k_i \approx \rho_i(t)v_{max}/k_i. \tag{11}$$

Then, the dynamic equation of densities on node $i$ is given by

$$l_i \frac{d\rho_i}{dt} = \sum_{j=1}^{k_i} \frac{v_{max}}{k_j}\rho_j(t) - v_{max}\rho_i(t). \tag{12}$$

Eq. (12) is the conventional diffusion equation with diffusion constant $v_{max}$ in network [33]. Thus, at low density, the dynamic equation of vehicular densities reduces the conventional diffusion equation on the cell-transmission graphs.

By taking $d\rho_i/dt = 0$, the equilibrium densities are obtained

$$\rho_{i,e} = \rho_{total}k_i / \sum_{j=1}^{N} k_j, \tag{13}$$

where $\rho_{total} = N\rho_0$, $\rho_{total}$ is the total density over the network, and $\rho_0$ is the mean density.

## 5. Analytical solutions for cell-transmission complete, cycle, and star graphs

We consider the traffic flow on the cell-transmission complete, cycle, and star graphs. We derive analytically the steady-state solutions of density equations on the cell-transmission complete, cycle, and star graphs.

We apply density equation (2) to the traffic flow on cell-transmission complete graph with node's number $N$. The density equations at node $i$ are given by

$$l_i \frac{d\rho_i}{dt} = \frac{1}{N-1} \sum_{j=1(j \neq i)}^{N} \rho_j u_i(\rho_i) - \frac{1}{N-1} \sum_{j=1(j \neq i)}^{N} \rho_i u_j(\rho_j). \tag{14}$$

At a steady state, all vehicular densities are the same because all nodes are topologically equivalent. The steady-state density at node $i$ is given by

$$\rho_{i,s} = \rho_0 = \rho_{total}/N, \tag{15}$$

where $\rho_0$ is the mean density, $\rho_{total}$ the total density, and $N$ the number of nodes.

The fundamental diagram at node (loop) $i$ is obtained from Eqs. (4) and (15)

$$J_{i,s} = v_{max}\rho_0 \quad \text{if } \rho_0 \leq 1/(v_{max}+1),$$
$$J_{i,s} = (1-\rho_0) \quad \text{if } \rho_0 > 1/(v_{max}+1). \tag{16}$$

We apply density equation (2) to the traffic flow on cell-transmission cycle graph with node's number $N$. The density equations at node $i$ is given by

$$l_i \frac{d\rho_i}{dt} = \rho_{i-1}u_i(\rho_i)/2 + \rho_{i+1}u_i(\rho_i)/2 - \rho_i u_{i-1}(\rho_{i-1})/2 - \rho_i u_{i+1}(\rho_{i+1})/2. \tag{17}$$

On the cell-transmission cycle graph, all nodes are topologically equivalent. Similarly to the cell-transmission complete graph, the steady-state density and fundamental diagram at node $i$ are obtained. They are given by Eqs. (15) and (16).

We apply density equation (2) to the traffic flow on cell-transmission star graph with node's number $N$. The density equations at node 1 (hub) with highest degree and node $i$ ($i \neq 1$) are given by

$$l_1 \frac{d\rho_1}{dt} = \sum_{j=2}^{N} \rho_j u_1(\rho_1) - \frac{1}{N-1}\sum_{j=2}^{N} \rho_1 u_j(\rho_j), \tag{18}$$

$$l_i \frac{d\rho_i}{dt} = \frac{1}{N-1} \rho_1 u_i(\rho_i) - \rho_i u_1(\rho_1). \tag{19}$$

Analytical solutions at a steady state can not be obtained for all values of density but one can obtain analytical solutions for low and high densities. At low density, the analytical solutions at a steady state are given from the diffusion approximation (13)

$$\rho_{1,s} = N\rho_0 / 2, \tag{20}$$

$$\rho_{i,s} = N\rho_0 / 2(N-1), \tag{21}$$

$$J_{1,s} = N v_{\max} \rho_0 / 2, \tag{22}$$

$$J_{i,s} = N v_{\max} \rho_0 / 2(N-1). \tag{23}$$

At high density, the density equations (18) and (19) are given by

$$l_1 \frac{d\rho_1}{dt} = \sum_{j=2}^{N} \rho_j (1-\rho_1)/\rho_1 - \frac{\rho_1}{N-1} \sum_{j=2}^{N} (1-\rho_j)/\rho_j, \tag{24}$$

$$l_i \frac{d\rho_i}{dt} = \frac{1}{N-1} \rho_1 (1-\rho_i)/\rho_i - \rho_i (1-\rho_1)/\rho_1. \tag{25}$$

At a steady state, the density at node $i$ ($i \neq 1$) is the same as those at other nodes except for hub 1 because the nodes except for node 1 are topologically equivalent. The density equations at a steady state reduce

$$(N-1)\rho_{2,s}(1-\rho_{1,s})/\rho_{1,s} - \rho_{1,s}(1-\rho_{2,s})/\rho_{2,s} = 0. \tag{26}$$

The following equation is obtained from the conservation law of densities

$$\rho_{1,s} + (N-1)\rho_{2,s} = N\rho_0. \tag{27}$$

We obtain the analytical solutions at a steady state from Eqs. (26) and (27). For any values of $N$, the expressions of steady-state solutions are very overlong. Here, we present the steady-state solutions for $N=3$ and $N=4$. The steady-state solutions for $N=3$ are given by

$$\rho_{1,s} = (-1+9\rho_0)/6 - (-1+54\rho_0 - 27\rho_0^2)/6C + C/6, \tag{28}$$

$$\rho_{2,s} = (1/6 + 3\rho_0/2 - 1/6C + 9\rho_0/C - 9\rho_0^2/2C - C/6)/2, \tag{29}$$

$$J_{1,s} = (1-\rho_{1,s}), \tag{30}$$

$$J_{2,s} = (1-\rho_{2,s}), \tag{31}$$

where

$$C = (-1+81\rho_0 -162\rho_0^2 +9\sqrt{3}\sqrt{-8\rho_0^2 +576\rho_0^3 -873\rho_0^4 +486\rho_0^5 -81\rho_0^6})^{1/3}.$$

The steady-state solutions for $N = 4$ are given by

$$\rho_{1,s} = (-1+6\rho_0)/3 - (-1+34\rho_0 -12\rho_0^2)/3C' + C'/3, \tag{32}$$

$$\rho_{2,s} = (1/3 + 2\rho_0 - 1/3C' + 8\rho_0/C' - 4\rho_0^2/C' - C'/3)/3, \tag{33}$$

$$J_{1,s} = (1-\rho_{1,s}), \tag{34}$$

$$J_{2,s} = (1-\rho_{2,s}), \tag{35}$$

where

$$C' = (-1+36\rho_0 -72\rho_0^2 +6\sqrt{3}\sqrt{-3\rho_0^2 +96\rho_0^3 -148\rho_0^4 +96\rho_0^5 -16\rho_0^6})^{1/3}.$$

## 6. Numerical results

We carry out numerical calculation for traffic flow on cell-transmission graphs. We calculate the vehicular densities and traffic currents on all roads for controlled traffic flow on star graphs. Hereafter, we set $l_i = 1$.

First, we calculate the vehicular densities and traffic currents on the graphs (b) and (d) shown in Fig. 1. The densities and currents at a steady state are obtained after sufficiently long time. Figure 3(a) shows the plots of steady-state densities $\rho_{i,s}$ ($i = 1,2,3$) against mean density $\rho_0$ for the traffic flow on cell-transmission complete graph with $N = 3$ in Fig. 1(b) where $v_{\max} = 5$ and $l_i = 1$ ($i = 1,2,3$). Black, red, green circles indicate the steady-state densities on nodes 1, 2, and 3 respectively. The solid line represents the analytical result (15). Densities on all nodes are the same. The simulation result agrees with the analytical result (15).

Figure 3(b) shows the plots of steady-state currents $J_{i,s}$ ($i = 1,2,3$) against mean density $\rho_0$ for the traffic flow on cell-transmission complete graph with $N=3$ in Fig. 1(b) where $v_{\max}=5$ and $l_i=1$ ($i=1,2,3$). Black, red, green circles indicate the steady-state currents on nodes 1, 2, and 3 respectively. The solid line represents the analytical result (16). Currents on all nodes are the same. The simulation result agrees with the analytical result (16).

Figure 4(a) shows the plots of steady-state densities $\rho_{i,s}$ ($i=1,2,3$) against mean density $\rho_0$ for the traffic flow on cell-transmission star graph with $N=3$ in Fig. 1(d) where $v_{\max}=5$ and $l_i=1$ ($i=1,2,3$). Black, red, green circles indicate the steady-state densities on nodes 1, 2, and 3 respectively. The steady-state density on node 3 is consistent with that on node 2. The steady-state density on hub 1 is higher than that on nodes 2 and 3. The linear lines with black and red colors represent the analytical results (20) and (21) at low mean density respectively. Eqs. (20) and (21) agree well with the simulation result at low density. The black and red curves indicate the analytical results (28) and (29) at high mean density respectively. Eqs. (28) and (29) are consistent with the simulation result at extended densities except for low mean density.

Figure 4(b) shows the plots of steady-state currents $J_{i,s}$ ($i=1,2,3$) against mean density $\rho_0$ for the traffic flow on cell-transmission star graph with $N=3$ in Fig. 1(d) where $v_{\max}=5$ and $l_i=1$ ($i=1,2,3$). Black, red, green circles indicate the steady-state currents on nodes 1, 2, and 3 respectively. The steady-state current on node 3 is consistent with that on node 2. The steady-state current on hub 1 is higher than that on nodes 2 and 3 at low mean density. While the steady-state current on hub 1 is lower than that on nodes 2 and 3 at immediate and high mean densities. The linear lines with black and red colors represent the analytical results (22) and (23) at low mean density respectively. Eqs. (22) and (23) agree well with the simulation result at low mean density. The black and red curves indicate the analytical results (30) and (31) at high mean density respectively. Eqs. (30) and (31) are consistent with the simulation result at extended densities except for low density.

We carry out numerical simulation for cell-transmission graphs with $N=4$ shown in Fig. 2. The cell-transmission complete and cycle graphs (a) and (b) are homogeneous. The values of densities and currents obtained by numerical simulation are the same on all nodes because all nodes are equivalent topologically. The plots of densities and currents against mean density agree with those in Fig. 3(a) and Fig. 3(b).

We calculate the vehicular densities and traffic currents on the cell-transmission star graph shown in Fig. 2(c). The densities and currents at a steady state are obtained after sufficiently long time. Figure 5(a) shows the plots of steady-state densities $\rho_{i,s}$ ($i=1,2,3,4$) against mean density $\rho_0$ for the traffic flow on cell-transmission star graph with $N=4$ in Fig. 2(c) where $v_{max}=5$ and $l_i=1$ ($i=1,2,3,4$). Black, red, green circles and dot indicate the steady-state densities on nodes 1, 2, 3, and 4 respectively. The steady-state density on nodes 3 and 4 is consistent with that on node 2. The steady-state density on hub 1 is higher than that on nodes 2, 3, and 4. The linear lines with black and red colors represent the analytical results (20) and (21) at low mean density respectively. Eqs. (20) and (21) agree well with the simulation result at low mean density. The black and red curves indicate the analytical results (32) and (33) at high mean density respectively. Eqs. (32) and (33) are consistent with the simulation result at extended densities except for low mean density. The difference between densities on hub 1 and other nodes is larger than that in Fig. 4(a) with $N=3$.

Figure 5(b) shows the plots of steady-state currents $J_{i,s}$ ($i=1,2,3,4$) against mean density $\rho_0$ for the traffic flow on cell-transmission star graph with $N=4$ in Fig. 2(c) where $v_{\max}=5$ and $l_i=1$ ($i=1,2,3,4$). Black, red, green circles, and dot indicate the steady-state currents on nodes 1, 2, 3, and 4 respectively. The steady-state current on nodes 3 and 4 is consistent with that on node 2. The steady-state current on hub 1 is higher than that on nodes 2, 3, and 4 at low mean density. While the steady-state current on hub 1 is lower than that on nodes 2, 3, and 4 at immediate and high densities. The linear lines with black and red colors represent the analytical results (22) and (23) at low mean density respectively. Eqs. (22) and (23) agree well with the simulation result at low mean density. The black and red curves indicate the analytical results (34) and (35) at high mean density respectively. Eqs. (34) and (35) are consistent with the simulation result at extended densities except for low mean density. Also, the difference between currents on hub 1 and other nodes is larger than that in Fig. 4(b) with $N=3$.

## 7. Summary

In the previous paper, we have presented the speed-matching model for vehicular traffic in the urban network [40]. In the model, we have proposed the expression $J_{i,j}(\rho_i,\rho_j)=f_{ij}\rho_i u_j(\rho_j)$ for the outflow from node $i$ to node $j$ where $f_{ij}$ is the fraction of branching flow from road $i$ to road $j$. The outflow depends on the velocity at the downstream. We have reproduced successfully the macroscopic fundamental diagrams for the urban-scale traffic on networks.

Here, we studied the traffic flow controlled on star graph analytically and numerically. We described the traffic star graph in terms of the cell-transmission graphs by using the cell-transmission method. We presented the dynamic equations of vehicular densities for the traffic flow on cell-transmission graphs. We found that the traffic flow on star graph is mapped to the nonlinear diffusion process on the cell-transmission graphs. We derived the analytical solutions for traffic flow at low and high densities on cell-transmission complete, cycle, and star graphs. We showed that the dynamic equations of densities are approximated by conventional diffusion equations at low density. We presented the numerical solutions and compared the numerical result with analytical result. We addressed how the urban-scale macroscopic fundamental diagram changes by the structure of the cell-transmission graph. The traffic flow controlled on star graph is successfully modeled by the cell-transmission graphs.

In this study, we studied the traffic flow controlled on star graph by using the cell-transmission graphs. In future, it will be necessary and important to extend the traffic model to the large-scale network. The analytical study about the traffic flow controlled on the star graph is the first and this study will be useful for the design and development of urban-scale traffic flow.

Figure captions

Fig. 1. (a) Schematic illustration of the star graph with $N=3$ where $N$ is the number of loops. Each loop represents one road. Loops 1, 2, and 3 are indicated by black, red, and blue colors. There is an intersection at the center. Vehicles move uni-directionally on each loop. After vehicles go round on a loop, they move out of the loop. After vehicles move out of loop 1, they are divided into two equal parts and flow into loops 2 and 3. Similarly, after they move out of loop 2, vehicles flow into loops 1 and 3 one half each. After vehicles move out of loop 3, they flow into loops 1 and 2 one half each. The process is repeated. The traffic flow from loop 1 (2, 3) to other loops is indicated by black (red, blue) arrow. (b) Cell-transmission graph transformed from traffic network (a). In the transformed graph, each node defines a road segment. Loops 1, 2, and 3 are represented by black, red, and blue nodes. A link represents the connectivity between roads. Traffic flow from node 1 (2, 3) to other nodes is represented by black (red, blue) arrow. The transformed graph is the cell-transmission complete graph with $N=3$. (c) Traffic flow different from that in network (a). After vehicles move out of loop 1, they are divided into two equal parts and flow into loops 2 and 3. After vehicles move out of loop 2, they flow only into loop 1. After vehicles move out of loop 3, they flow only into loop 1. The process is repeated. The traffic flow from loop 1 (2, 3) to other loops is indicated by black (red, blue) arrow. (d) Cell-transmission graph transformed from traffic network (c). Traffic flow from node 1 (2, 3) to other nodes is represented by black (red, blue) arrow. The transformed graph is the cell-transmission star graph with $N=3$. There is no links between nodes 2 and 3.

Fig. 2. All cell-transmission graphs for traffic star graph with $N=4$. In the cell-transmission graphs, indegree $k_{in}$ is consistent with outdegree $k_{out}$, $k_{in} = k_{out}$.

Fig. 3. (a) Plots of steady-state densities $\rho_{i,s}$ ($i=1,2,3$) against mean density $\rho_0$ for the traffic flow on cell-transmission complete graph with $N=3$ in Fig. 1(b) where $v_{max} = 5$ and $l_i = 1$ ($i=1,2,3$). Black, red, green circles indicate the steady-state densities on nodes 1, 2, and 3 respectively. The solid line represents the analytical result (15). Densities on all nodes are the same. (b) Plots of steady-state currents $J_{i,s}$ ($i=1,2,3$) against mean density $\rho_0$ for the traffic flow on cell-transmission complete graph with $N=3$ in Fig. 1(b) where $v_{max} = 5$ and $l_i = 1$ ($i=1,2,3$). Black, red, green circles indicate the steady-state currents on nodes 1, 2, and 3 respectively. The solid line represents the analytical result (16). Currents on all nodes are the same.

Fig. 4. (a) Plots of steady-state densities $\rho_{i,s}$ ($i=1,2,3$) against mean density $\rho_0$ for the traffic flow on cell-transmission star graph with $N=3$ in Fig. 1(d) where $v_{max} = 5$ and $l_i = 1$ ($i=1,2,3$). Black, red, green circles indicate the steady-state densities on nodes 1, 2, and 3 respectively. The steady-state density on node 3 is consistent with that on node 2. The linear lines with black and red colors represent the analytical results (20) and (21) at low density respectively. The black and red curves indicate the analytical results (28) and (29) respectively. (b) Plots of steady-state currents $J_{i,s}$ ($i=1,2,3$) against mean density $\rho_0$ for the traffic flow on cell-transmission star graph with $N=3$ in Fig. 1(d) where $v_{max} = 5$ and $l_i = 1$ ($i=1,2,3$). Black, red, green circles indicate the steady-state currents on nodes 1, 2, and 3 respectively. The steady-state current on node 3 is consistent with that on node 2. The linear lines with black and red colors represent the analytical results (22) and (23) at low density respectively. The black and red curves

indicate the analytical results (30) and (31) respectively.

Fig. 5. Plots of steady-state densities $\rho_{i,s}$ ($i = 1,2,3,4$) against mean density $\rho_0$ for the traffic flow on cell-transmission star graph with $N = 4$ in Fig. 2(c) where $v_{max} = 5$ and $l_i = 1$ ($i = 1,2,3,4$). Black, red, green circles and dot indicate the steady-state densities on nodes 1, 2, 3, and 4 respectively. The steady-state density on nodes 3 and 4 is consistent with that on node 2. The linear lines with black and red colors represent the analytical results (20) and (21) at low mean density respectively. The black and red curves indicate the analytical results (32) and (33) at high mean density respectively. (b) Plots of steady-state currents $J_{i,s}$ ($i = 1,2,3,4$) against mean density $\rho_0$ for the traffic flow on cell-transmission star graph with $N = 4$ in Fig. 2(c) where $v_{max} = 5$ and $l_i = 1$ ($i = 1,2,3,4$). Black, red, green circles, and dot indicate the steady-state currents on nodes 1, 2, 3, and 4 respectively. The steady-state current on nodes 3 and 4 is consistent with that on node 2. The linear lines with black and red colors represent the analytical results (22) and (23) at low mean density respectively. The black and red curves indicate the analytical results (34) and (35) at high mean density respectively.

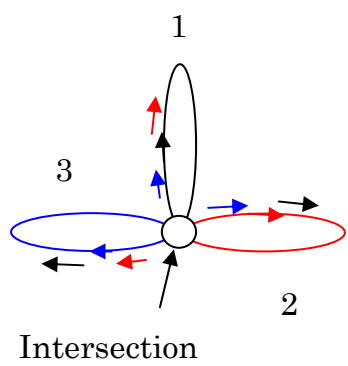
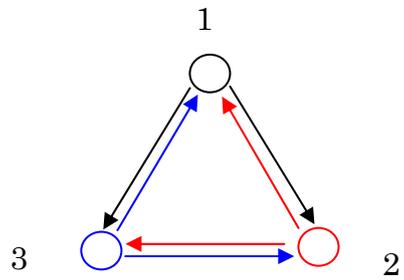

(a)              (b)

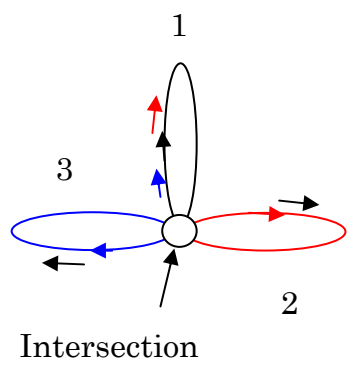
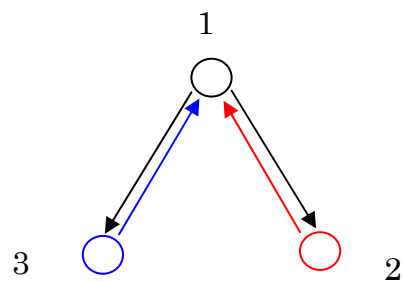

(c)              (d)

Fig. 1.

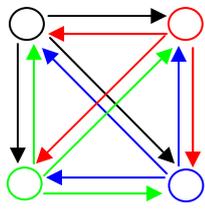 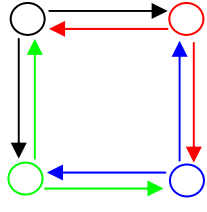 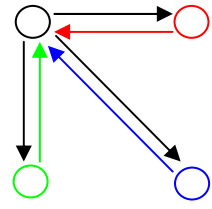

(a) (b) (c)

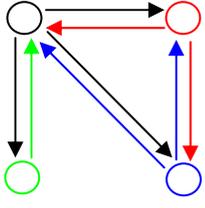 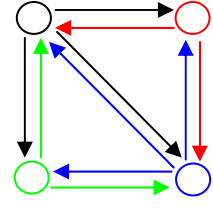 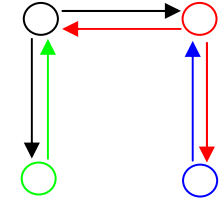

(d) (e) (f)

Fig. 2.

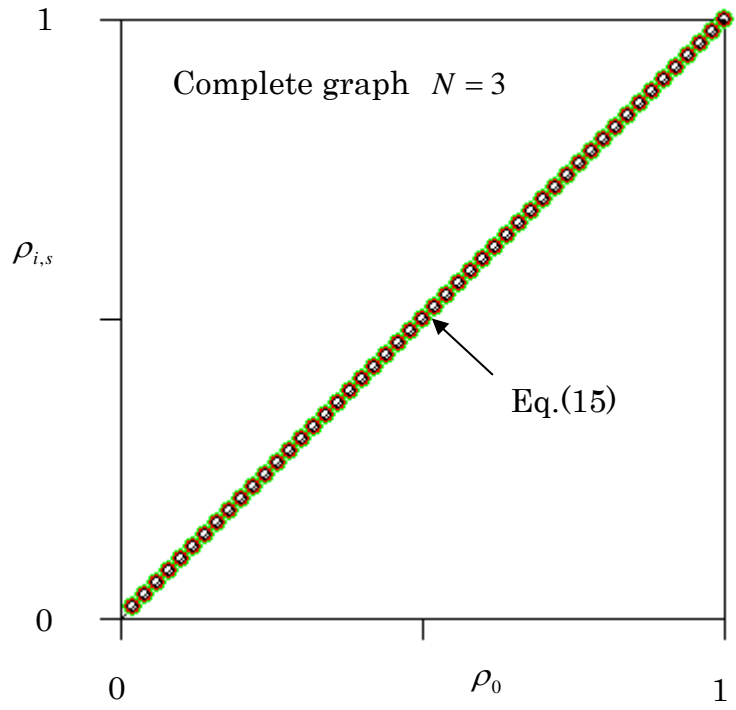

Fig. 3(a).

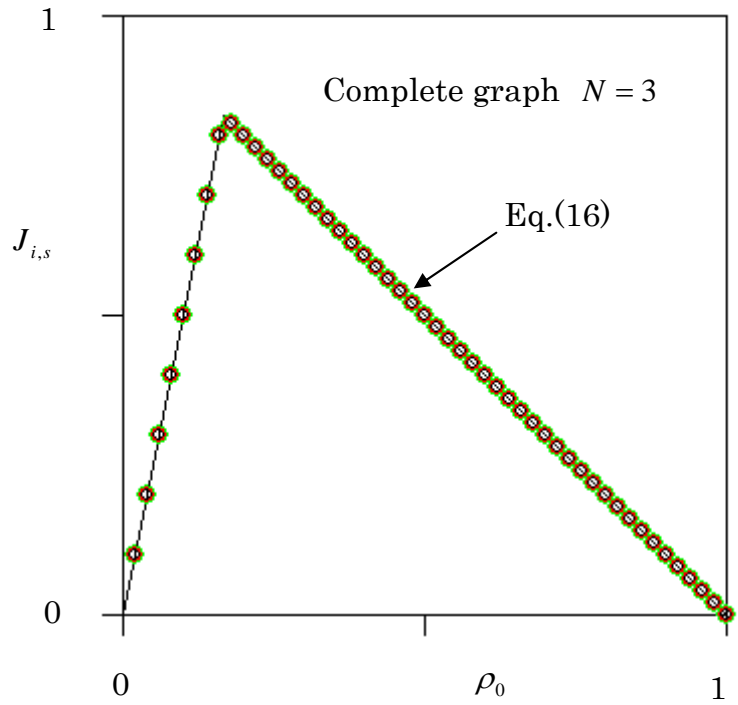

Fig. 3(b).

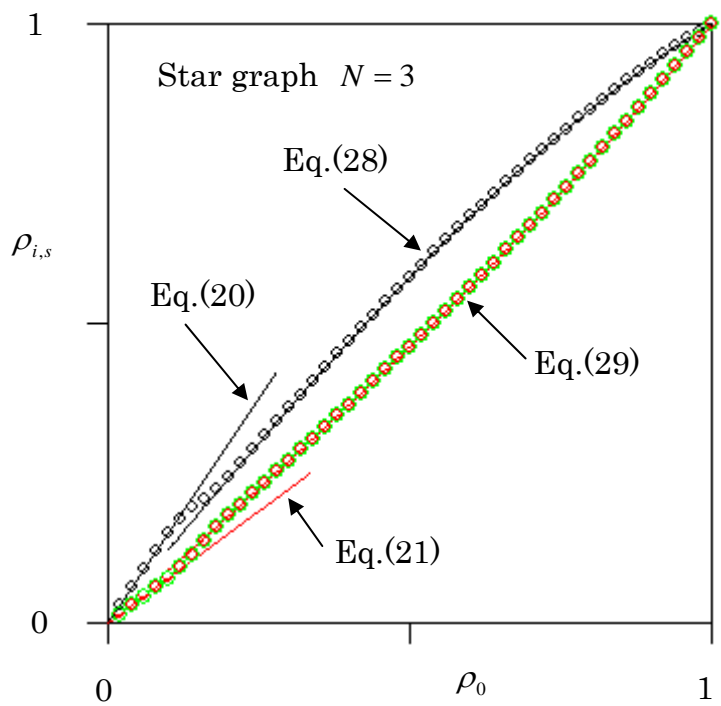

Fig. 4(a).

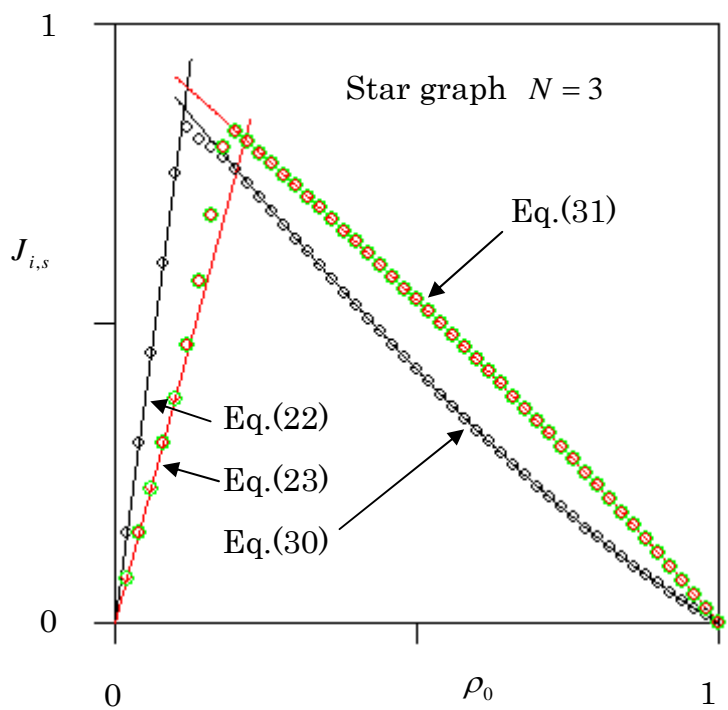

Fig. 4(b).

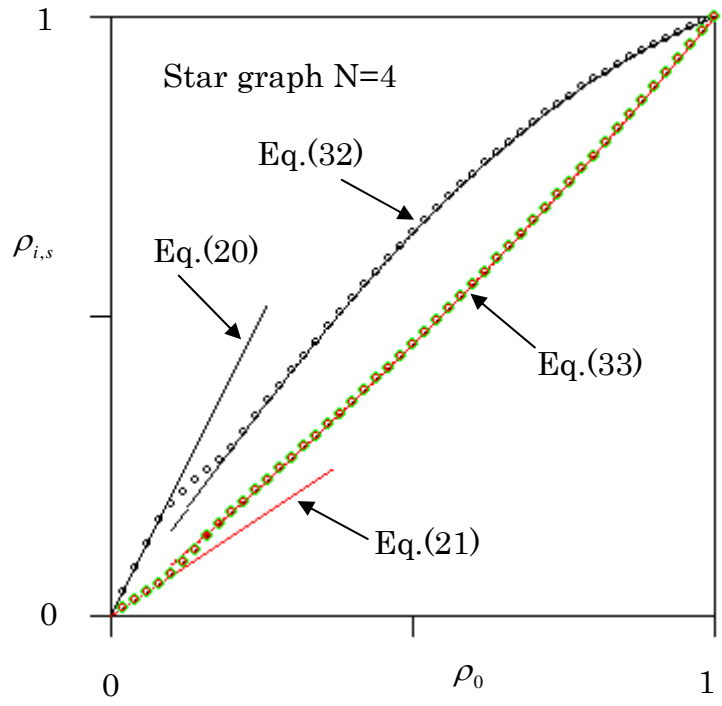

Fig. 5(a).

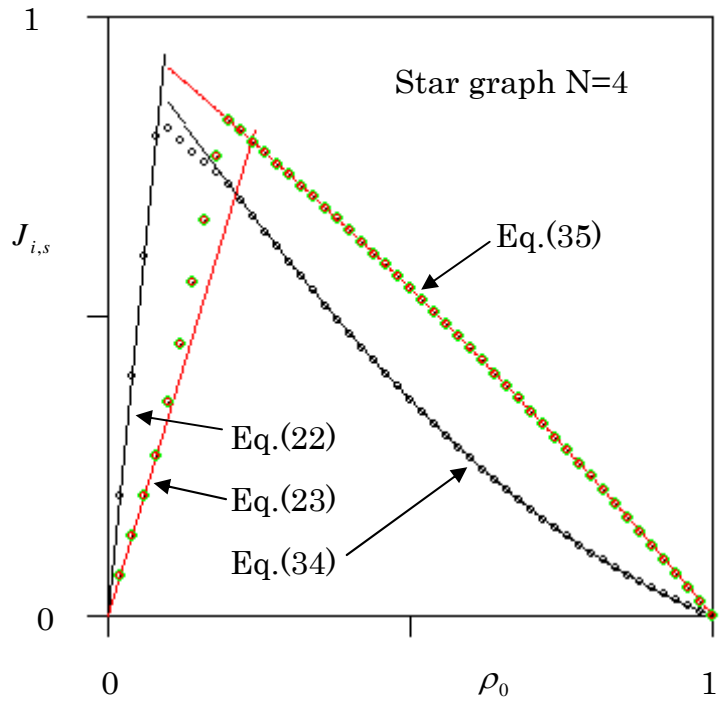

Fig. 5(b).